\documentclass[%
reprint,
showpacs,preprintnumbers,
 amsmath,amssymb,
 aip,
 graphicx,
apl,
]{revtex4-1}

\usepackage{graphicx}
\usepackage{dcolumn}
\usepackage{bm}
\usepackage[mathlines]{lineno}

\newcommand{\gst}{\ensuremath{\mathrm{Ge_{2}Sb_{2}Te_{5}}}}

\newcommand{\getef}{\ensuremath{\mathrm{GeTe_{4}}}}
\newcommand{\getes}{\ensuremath{\mathrm{GeTe_{6}}}}

\newcommand{\mr}[1]{\ensuremath{\mathrm{#1}}}

\begin{document}
\preprint{} 

\title[Short Title]{Polarization Dependent Optical Control of Atomic Arrangement in Multilayer Ge-Sb-Te Phase Change Materials}

\author{Kotaro Makino}
\affiliation{Institute of Applied Physics, University of Tsukuba, 1-1-1 Tennodai, Tsukuba 305-8573, Japan}
\author{Junji Tominaga}
\author{Alexander V. Kolobov}
\author{Paul Fons}
\affiliation{Nanoelectronics Research Institute, National Institute of Advanced Industrial Science and Technology, Tsukuba Central 4, 1-1-1 Higashi, Tsukuba 305-8562, Japan}
\author{Muneaki Hase}
\affiliation{Institute of Applied Physics, University of Tsukuba, 1-1-1 Tennodai, Tsukuba 305-8573, Japan}

\date{\today}

\begin{abstract}
We report the optical perturbation of atomic arrangement in  the layered GeTe/Sb$_{2}$Te$_{3}$ phase change memory material.
To observe the structural change, the coherent A$_1$ mode of GeTe$_4$ local structure is investigated at various polarization angles of femtosecond pump pulses with the fluence at $\leq$ 78 $\mu$J/cm{$^2$}. $p$-polarization found to be more effective in inducing the A$_1$ frequency shift that can be either reversible or irreversible, depending on the pump fluence. The predominant origin of this shift is attributed to rearrangement of Ge atoms driven by anisotropic dissociation of the Ge-Te bonds along the [111] axis after the $p$-polarized pulse irradiation.
\end{abstract}


\maketitle
 
The irradiation of intense femtosecond laser pulses into 
opaque materials has led to ultrafast phase changes due to the photo-excitation of electrons from bonding into anti-bonding states \cite{Solis96,Tinten98,Callan01}. This phenomenon has widely been accepted as an electronic excitation induced phase transition (or electronic melting) due to strong perturbation of bonding nature, which has also been extensively reported in a variety of narrow band-gap semiconductors, such as Si \cite{Shank83}, GaAs \cite{Huang98}, Ti$_{2}$O$_{3}$ \cite{Cheng93}, carbon nanotubes \cite{Traian06} and of semimetals, such as Bi \cite{Hase02} and graphite \cite{Ishioka08}.
Most of the experimental and theoretical studies have been carried out under the irradiation of amplified femtosecond laser pulses, whose fluences of more than several mJ/cm$^2$.

Electrical and optical phase-change memory technologies utilizing phase change between amorphous (disordered) and crystalline (ordered) phases consisted of \gst (GST) have been extensively used in commercial applications, such as DVD, Blu-lay disk, and electrical memory \cite{Wuttig07}. 
One of the benefits of using GST as such the applications is 
more than 10 \% contrast in optical (reflectivity) and 10$^{3}$ contrast in electronic (resistivity) 
properties upon the transformation between the disordered and ordered phases induced by lattice heating with nanosecond optical or electrical pulses \cite{Wuttig07}. 
However, the physical origin of the high contrast in optical and electrical properties upon the phase change has long been debated \cite{Kolobov04}. 
A recent ellipsometric study revealed the significant difference in the electrical and optical properties originated from resonant bonding, \cite{Pauling39} realized by long-range atomic order \cite{Shportko07}.
In GST materials, therefore, ultrafast laser perturbation of atomic arrangement presuming on bond energy hierarchy \cite{Kolobov11} is expected to be a possible method for ultrafast switching between the ordered and disordered phases.

In the last decade, the phase change in GST materials induced by pico- or femtosecond laser pulse irradiation has been extensively investigated\cite{Siegel04, Siegel08,Zhang06}, and suggested that threshold fluence for the phase change between the crystalline and amorphous phases was several tens of mJ/cm$^2$.
Since during such the laser-induced phase change in GST, lattice heating generally happens, in which most part of the energy absorbed is lost due to thermal diffusion, making such the high threshold fluence.
Limited studies have been reported on the low threshold (e.g., $\sim \mu$J/cm$^2$) phase change in GST systems. 

In this Letter, we present a polarization dependent optical perturbation of atomic bonding in multilayer GeTe/Sb$_{2}$Te$_{3}$ superlattice (SL) thin films using a low fluence pump-probe technique. 
Recently, lower current switching of the multilayer Ge-Sb-Te phase change memory (PCM) has been reported \cite{Simpson11}. 
Due to the low phase change switching energy, it would be possible to reduce the fluence of femtosecond pulse when we 
consider the optical perturbation of atomic arrangement in the GeTe/Sb$_{2}$Te$_{3}$ memory materials.
In fact, we have recently reported the low fluence and ultrafast phase switching in GeTe/Sb$_{2}$Te$_{3}$ SL using double pump pulses,\cite{Makino11} in which the main player was the selectively excited coherent vibrational motion of the local mode of \getef. 
On the other hand, significant dependence of polarization angle of pump pulse is expected because of the SL structure\cite{Chiu11}.  
As we expected, by precisely tuning the polarization angle of the pump beam at $\leq$ 78 $\mu$J/cm$^2$, the frequency of a local vibrational motion observed in GeTe/Sb$_{2}$Te$_{3}$ SL significantly red-shifts from the characteristic value in the disordered phase toward that of the ordered phase. 
We attribute the polarization dependent red-shift to the modification in equilibrium position of Ge atoms resulting from anisotropic photo-excited carriers. 

\begin{figure}[htbp]
\includegraphics[width=85mm]{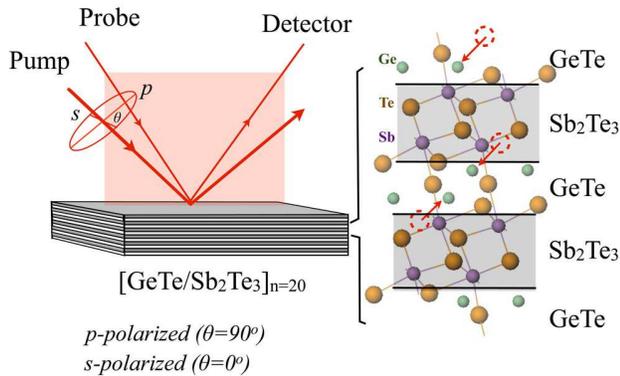}
\caption {(color online) Schematic of the polarization-dependent pump-probe experiment. $\theta$ represents the angle from the surface parallel; $\theta$ = 0{$^\circ $} means the pump beam is $s$-polarized, while $\theta$ = 90{$^\circ $} means $p$-polarized. Schematic of the structure of GeTe/Sb$_{2}$Te$_{3}$ SL is shown in the magnified view. Ge atoms are green, Te atoms are tan and Sb atoms are purple.}
\label{Fig. 1}
\end{figure}

Figure 1 illustrates the schematic view of our experimental setup and structure of the sample. 
The samples used in the present study were thin films of disordered (amorphous) GeTe/Sb$_{2}$Te$_{3}$ SL fabricated using a helicon-wave RF magnetron sputtering machine on Si (100) substrate. The thickness of the films was 20 nm \cite{Tominaga08}.
Time-resolved pump-probe reflectivity measurements were performed utilizing a near-infrared optical pulse with 20 fs duration and a center wavelength of 850~nm generated by a Ti:sapphire laser oscillator. The linearly polarized pump and probe pulses were focused onto the sample surface by a 33.85~mm focal length 60{$^\circ $}-off-axis parabolic mirror.
The excitation of the GeTe/Sb$_{2}$Te$_{3}$ SLs  with the 850 nm (= 1.46 eV) laser pulse generates photo-carriers across the narrow band gap of $\approx$ 0.5 - 0.7 eV \cite{Lee05}. 
{The average power of the pump beams was 3 or 15 mW while that of probe beam was fixed at 0.5 mW, and we estimated the pump fluence to be 16 or 78 $\mu$J/cm$^{2}$ and the probe fluence to be 3 $\mu$J/cm$^2$, respectively.
The transient reflectivity change of the probe pulse ($\Delta R/R$) was recorded as a function of the pump-probe time delay ($ \tau $) at room temperature. 
As derived from the two temperature model (TTM)\cite{Allen87}, in which the heat capacities of the electron and lattice sub-systems are included, the pump fluence of 78 $\mu$J/cm$^2$ would induce the increase in the lattice temperature of only $\leq$ 10 K, and thus it signifies that accumulative thermal effect should be negligibly small, and the excitation is below the threshold of thermal crystallization\cite{Simpson11}.

In our sample, as shown in the right hand side of Fig. 1, the phase change is dominantly characterized by the displacement of Ge atoms at the angle of 45{$^\circ $} to the sample normal ([111] direction of the local structure), and therefore, oblique incident $p$-polarized light can effectively excite the atomic bonds along the [111] direction \cite{Pfeifer92}.
The incident angles of the pump and probe pulses were $ \approx $ 50{$^\circ $} and $ \approx $ 40{$^\circ $} from the surface normal, respectively. The polarization angle of the pump pulse ($ \theta$) was varied from 0{$^\circ $} to 90{$^\circ $} ($p$-polarization) and then changed to 180{$^\circ $} ($s$-polarization) 
by manually rotating the half-wave plate with every several minutes of the data scan, while that of the probe pulse was kept at 90{$^\circ $}. 
The amplitude of the coherent lattice displacement (modulation of Ge-Te bond length) estimated from the maximum excited carrier density of 5.3$\times$10$^{19}$cm$^{-3}$ is $\sim$ 2.6$\times$10$^{-4}$ nm,\cite{Zeiger92} which is only $\approx$ 10$^{-3}$ of the Ge--Te bond length found in GST alloy (0.26 nm). 
This signifies that there is no anharmonic phonon softening in the present condition\cite{Hase02}.

\begin{figure}[htbp]
\includegraphics[width=88mm]{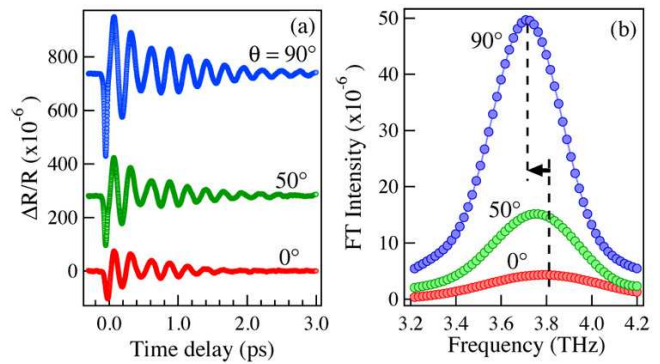}
\caption {(color online) (a) Time-resolved $ \Delta R/R$ signals and (b) the corresponding FT spectra measured with the pump fluence of $\approx$ 78 $\mu$J/cm{$^2$} and $ \theta = 0^\circ, 50^\circ $ and $ 90^\circ $. 
Here, $\theta = $ 0{$^\circ $} corresponds to the $s$-polarization, while $ \theta = $90{$^\circ $} corresponds to the $p$-polarization. Dashed lines indicate central frequencies of peaks for $ \theta = 0^\circ $ and $ \theta = 90^\circ $, respectively.
}
\label{Fig. 2}
\end{figure}

Figure 2(a) summarizes the transient reflectivity change ($ \Delta R/R$) for three different $ \theta$s as a function of $ \tau $ for GeTe/Sb$_{2}$Te$_{3}$ SL films measured with the pump fluence of $ \approx $78 $ \mu $J/cm{$^2$}. Just after $\tau $ = 0, coherent oscillations appear, which dominate over 
the electronic response due to the generation of photo-excited carriers. The Fourier transformed (FT) spectra obtained from Fig. 2(a) are displayed in Fig. 2(b), in which the local \mr{A_1} mode of \getef~at $\approx$ 3.8 THz is mainly observed and focused in the present study\cite{Forst00,Hase09,Rueda11}. 
By increasing $ \theta$ from 0{$^\circ $} to 90{$^\circ $}, a red-shift as well as the enhancement of the intensity of the \mr{A_1} mode of \getef~ is observed. 
By using the formula of Fresnel reflection, we confirmed that the enhancement in the phonon intensity is dominantly a consequence of the difference 
in reflectivity when varying the polarization angle at the sample surface. With respect to the red-shift of the A$_{1}$ mode (from $\approx$ 3.80 to $ \approx$ 3.70 THz), however, we cannot attribute this shift of $\approx$ 0.1 THz to the simple phonon softening because we previously reported that the red-shift was only 0.04 THz even when excitation fluence was increased from 47 to 284 $\mu$J/cm{$^2$} with tenfold increase in phonon amplitude \cite{Makino11}. Therefore, it is supposed that the red-shift observed in Fig. 2(b) originates from the rearrangement of atoms, e.g., the change in equilibrium position of Ge (and surrounding atoms), as discussed later.

\begin{figure}[htbp]
\includegraphics[width=70mm]{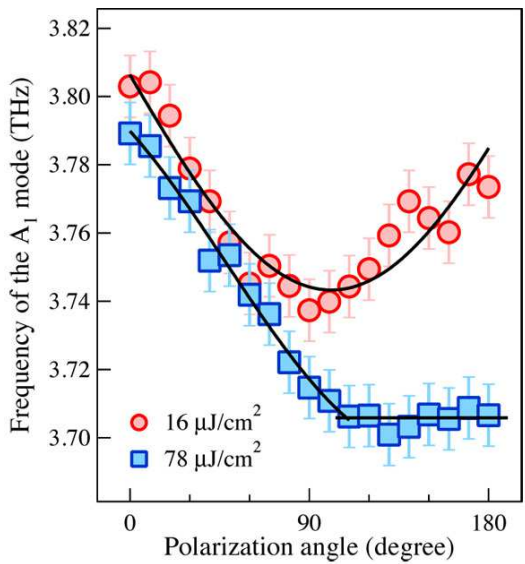}
\caption {(color online) The polarization dependence of the frequency of the coherent local A{$_1$} mode of \getef~measured at 0 
$ \leqq $  $ \theta $ $ \leqq $ 180{$^\circ $}, with the pump fluence of 16 and 78 $\mu$J/cm{$^2$}. The solid black lines are guides for the eyes.}
\label{Fig. 3}
\end{figure}

To obtain deeper insights into the nature of the observed $ \theta$-dependence of the \mr{A_1} mode frequency, the pump polarization angle dependence of the center frequency of the coherent local A{$_1$} mode of GeTe{$_4$} is displayed in Fig. 3.
Here, the center frequencies are extracted from the FT spectra by fitting to a Lorenz function with the best accuracy of 0.01 THz. Although Raman measurements have observed phonon spectra of GeTe\cite{Salicio09} and GST\cite{Andrikopoulos07}, the linewidth of the \mr{A_1} mode in Raman spectra was broader and the frequency resolution would not be good enough to resolve the 0.01 THz shift. Thus, more accurate analysis on the frequency shift in the relatively lower frequency region was possible in the coherent phonon spectroscopy, as demonstrated by Ishioka {\it et al} \cite{Ishioka08}. 
At the lower limit of the pump fluence (16 $\mu$J/cm{$^2$}), the A{$_1$} mode frequency is softened from $ \approx$ 3.80 to $ \approx$ 3.74 THz when $ \theta$ is changed from 0{$^\circ $} to 90{$^\circ $}, followed by the hardening of the A{$_1$} mode frequency when $ \theta$ is changed from 90{$^\circ $} to 180{$^\circ $}.
At the highest pump fluence (78 $\mu$J/cm{$^2$}), on the other hand, it is softened from $ \approx$ 3.80 to $ \approx$ 3.70 THz when $ \theta$ is changed from 0{$^\circ $} to 90{$^\circ $}, but never comes back to the original value even when $ \theta$ is changed to 180{$^\circ $}, suggesting that excitation with the $p$-polarized pump pulse induces both reversible and irreversible changes of phonon frequency, which is controlled by the pump fluence. 

The phonon relaxation time obtained by the fitting of the time-domain data in Fig. 2(a) with a damped harmonic oscillation increases from 0.68 to 1.04 ps as the polarization angle is rotated from 0{$^\circ $} to 90{$^\circ $} (not shown). 
Given the fact that lattice heating generally induces the red-shift of the frequency and decrease in the relaxation time for coherent optical phonons \cite{Hase98}, the observed effects that the phonon frequency decreases, while the relaxation time increases 
cannot simply be explained by lattice heating by the pump pulse. 
Instead, optical perturbation of local atomic arrangements would be the main origin of the observation in Fig. 3. 
Because the lowest value of $\approx$ 3.70 THz corresponds to the A{$_1$} mode frequency in the ordered phase \cite{Hase09, Makino11}, the highest pump fluence (78 $\mu$J/cm{$^2$}) with  $p$-polarization could induce atomic rearrangement in GeTe/Sb$_{2}$Te$_{3}$ SL  from tetrahedral-like (\getef)~ to octahedral-like (\getes)~coordinations, if we refer the umbrella flip model as the dominant mechanism \cite{Kolobov04}. 
Note that, a series of experimental data with 0{$^\circ $} $ \leqq $ ($ \theta$) $ \leqq $ 180{$^\circ $} was taken by multi pulse shots on the same irradiated exposure area.
Thereafter, the irreversible phase change observed with higher fluence may be influenced by the accumulation of laser shots.
This should be a subject of further study, by taking the rate of the phase change per laser shot into consideration \cite{Kanasaki09}. A single shot measurement using lower repetition rate femtosecond laser will be useful for that purpose.

We discuss why the $p$-polarized pump pulse induced local atomic arrangement based on the phase change model. 
The $p$-polarized pulse excitation would generate stronger anisotropic photo-excited carrier distribution in the [111] direction, along 
which weaker Ge-Te bonds are aligned, than in the case of the $s$-polarized pulse excitation \cite{Pfeifer92}. 
Here, we assume that weaker Ge-Te bonds are the bonds that are broken during the displacement of Ge atoms due to the bond energy hierarchy. 
After that, the anisotropic carrier-phonon interaction \cite{Carbone08} (via e.g., deformation potential) along the Ge-Te line results in the displacement of Ge atoms from tetrahedral site toward octahedral site. 
The estimated atomic displacement of the 2.6$\times$10$^{-4}$ nm ($\approx$ 10$^{-3}$ of the Ge--Te bond length ) seems not large enough to induce the observed phase change, thereby, the displacement should be enhanced when exciting with $p$-polarized pulse. 
Note that another pathway \cite{Huang10} of phase change in GeTe/Sb$_{2}$Te$_{3}$ SL was also presented in Ref \cite{Simpson11}. In short, the phase change was characterized as small displacement of Ge atoms along [110] direction. However, this pathway cannot explain our results in Figs. 2 and 3 well.

To check if our observation is not a general phenomenon, we measured the polarization dependence of the A{$_1$} mode frequency in the amorphous GST alloy films and found that the polarization dependence of the A{$_1$} mode frequency was not clearly resolved, although the small enhancement of the A{$_1$} phonon amplitude was observed as in the case of GeTe/Sb$_{2}$Te$_{3}$ SL. 
Also, the coherent  A{$_{1g}$} mode of Bi (0001) single crystal was measured under the same condition, and confirmed that there was no frequency red-shift when the pump polarization was varied. In fact, the pump fluence presented here is so low that phonon softening would not be observed in bulk semimetals \cite{Hase02}. 

\begin{figure}[htbp]
\includegraphics[width=80mm]{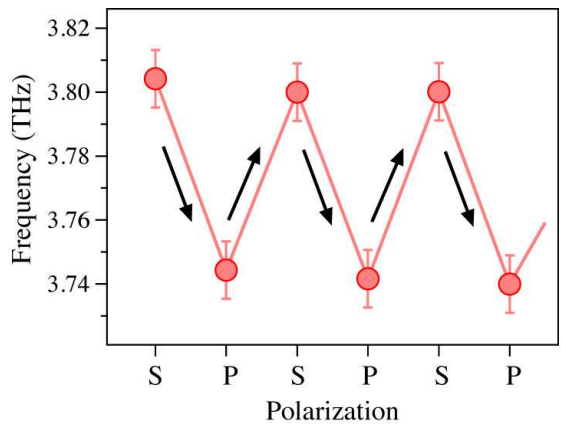}
\caption {(color online) Frequencies of the coherent local A$_{1}$ mode obtained by the sequence of the $p$-polarized and $s$-polarized pump pulse at 16 $\mu$J/cm{$^2$}, which demonstrate the repetitive switching between the two phases. The horizontal axis indicates the polarization states.}
\label{Fig. 4}
\end{figure}

Finally, we demonstrate the phase switching by repetitively changing the polarization angle of the pump pulse below the threshold of the irreversible phase change (16 $\mu$J/cm{$^2$}) as shown in Fig. 4. The frequency of the local A{$_1$} mode follows switching between $\approx$3.80 THz ($p$-polarized) and $\approx$3.74 THz ($s$-polarized). 
This means that polarization-induced modulation of the equilibrium position of Ge atoms can be achieved by using linearly polarized optical pulse irradiation. It is to be noted that the phase characterized by the 3.74 THz frequency would be a kind of long-lived intermediate ('hidden') phase between the disordered and ordered phases\cite{Ichikawa11}, which can be observed only when using electronic excitation\cite{Kolobov11,Makino11}.
The detailed dynamics of the structure of the ÔhiddenÕ  phase in GST superlattice will be observed by using a femtosecond x-ray diffraction measurements, which is beyond the scope of the present paper, but we will try to do such experiment in the future.

In conclusion, we have demonstrated bond-selective excitation in GeTe/Sb$_{2}$Te$_{3}$ SL by linearly polarized optical pulse excitation. 
The observation of coherent local vibrational motion has been carried out with various pump pulse polarization. 
When the polarization of the pump pulse was purely $p$-polarization, not only the increase in the phonon intensity but also the decrease in the phonon frequency, which can be attributed to the rearrangement of Ge atoms, were observed, over the sufficiently lower pump fluence range below 78 $\mu$J/cm{$^2$}. 
We found that reversible and irreversible phase change can be introduced in GeTe/Sb$_{2}$Te$_{3}$ SL 
depending on the pump fluence. The results obtained here will be important for the application of GeTe/Sb$_{2}$Te$_{3}$ materials as ultrafast and power consumption optical data storage devices. 

This work was supported in part by KAKENHI-22340076 from MEXT, Japan. 
K. M. acknowledges JSPS research fellowship from JSPS, Japan.

\end{document}